# The effect of primary treatment of wastewater in high rate algal pond systems: biomass and bioenergy recovery

Larissa T. Arashiro[a,b], Ivet Ferrer[a]*, Diederik P.L. Rousseau[b], Stijn W.H. Van Hulle[b] and Marianna Garfí[a]

[a]GEMMA - Group of Environmental Engineering and Microbiology, Department of Civil and Environmental Engineering, Universitat Politècnica de Catalunya · BarcelonaTech, c/ Jordi Girona 1-3, Building D1, 08034 Barcelona, Spain

[b]Department of Green Chemistry and Technology, Ghent University Campus Kortrijk, Graaf Karel de Goedelaan 5, 8500 Kortrijk, Belgium

* Corresponding author: Tel: +34 934016463
*E-mail address*: ivet.ferrer@upc.edu (I. Ferrer)

**Abstract**

The aim of this study was to assess the effect of primary treatment on the performance of two pilot-scale high rate algal ponds (HRAPs) treating urban wastewater, considering their treatment efficiency, biomass productivity, characteristics and biogas production potential. Results indicated that the primary treatment did not significantly affect the wastewater treatment efficiency ($NH_4^+$-N removal of 93 and 91% and COD removal of 62 and 65% in HRAP with and without primary treatment, respectively). The HRAP without primary treatment had higher biodiversity and productivity (18 vs. 16 g VSS/$m^2$d). Biomass from both systems presented good settling capacity. Results of biochemical methane potential test showed that co-digesting microalgae and primary sludge led to higher methane yields (238 - 258 mL $CH_4$/g VS) compared with microalgae mono-digestion (189 - 225 mL $CH_4$/g VS). Overall, HRAPs with and without primary treatment seem to be appropriate alternatives for combining wastewater treatment and bioenergy recovery.

# 1. Introduction

High rate algal ponds (HRAPs) have received renewed interest due to their capacity to treat wastewater with reduced energy consumption compared to conventional activated sludge systems, while producing microalgal biomass that can be used for non-food bioproducts and biofuels production (Young et al., 2017). HRAPs consist of shallow, paddlewheel mixed, raceway ponds where microalgae assimilate nutrients and produce oxygen, which is used by bacteria to oxidise organic matter (Craggs et al., 2014; Park et al., 2011). They are low-cost technologies that can be successfully implemented in locations where weather conditions are favourable for microalgae growth (e.g. high solar radiation and temperature). These natural systems are appropriate solutions for wastewater treatment especially in small agglomerations, since they reduce costs and environmental impacts associated with wastewater treatment (Garfí et al., 2017). In this context, they were reported to treat anaerobically digested domestic wastewater reaching removal efficiencies of up to 97% of $NH_4^+$-N and 87% of soluble biochemical oxygen demand ($sBOD_5$) at optimal conditions (Park and Craggs, 2011). Similar results were obtained from HRAPs treating primary settled urban wastewater, reaching average removal of 80% of chemical oxygen demand (COD) and 95% of $NH_4^+$-N (Gutiérrez et al., 2016). Other studies applying this technology to treat agricultural wastes and industrial wastewater were also reported (de Godos et al., 2010; Ibekwe et al., 2017; Van Den Hende et al., 2016). Moreover, HRAPs have been proven to be very effective for the recovery of bioenergy (e.g. biofuels), nutrients (e.g. biofertilisers) and valuable compounds (e.g. pigments, lipids) from wastewater (Arashiro et al., 2018; Craggs et al., 2011; Van Den Hende et al., 2016).

The installation and maintenance of HRAPs are significantly cheaper compared to conventional activated sludge systems and closed photobioreactors (Delrue et al., 2016). Another advantage of the HRAPs is that greenhouse gas emissions are also reduced, making them an option to improve the sustainability of wastewater treatment (Acién et al., 2016). However, one of the main drawbacks for implementing HRAPs for wastewater treatment is the large surface area requirement (up to 6 $m_2$/PE), which is necessary to promote satisfactory removal efficiency and biomass productivity. Indeed, a critical analysis of the latest studies on microalgae-based processes for wastewater treatment identified that the major obstacle hindering the dissemination of these technologies is the land requirement (Acién et al., 2016). In order to overcome this drawback and to simplify system operation and maintenance, the option of removing the primary treatment from the entire process could be considered. Primary treatment consists of removing settleable organic and inorganic solids from the raw wastewater by sedimentation. To date, there are several studies on optimising the HRAP operating conditions, such as depth, hydraulic retention time (HRT) and dynamics (Amini et al., 2016; Buchanan et al., 2018; Sutherland et al., 2014). However, there are no studies in the literature which investigate, in practice, the role and effect of the primary treatment step before the HRAPs. Posadas et al. (2017) carried out a theoretical case study suggesting that primary suspended solids removal is probably unnecessary in a HRAP system. This implication was based on the fact that the removal of biodegradable suspended solids can be efficiently reached by microalgal photosynthesis, which generates large excess in oxygenation capacity in the ponds. As suspended solids from raw wastewater may have an impact on light penetration and microalgae growth, which is directly related to biomass productivity and treatment capacity, further research is needed in order to demonstrate the feasibility of this configuration. Moreover, the possibility of incorporating a downstream process for microalgae biomass valorisation could be jeopardised in case the quality and amount of biomass was negatively affected by the absence of primary treatment.

Facing the current energy and environmental crisis, with the global economy relying on fossil fuels, extensive research has been done to valorise microalgal biomass within a biorefinery approach (Raheem et al., 2018; Šoštarič et al., 2012). Among the different biomass valorisation techniques proposed so far, biogas production seems to be the least complex option to recover bioenergy from microalgal biomass. Previous studies have reported the microalgae as a potential substrate for anaerobic digestion, especially after undergoing pretreatments to enhance the methane yield (González-Fernández et al., 2012; Uggetti et al., 2017).

The aim of this research was therefore to investigate the effect of primary treatment on the long-term performance of pilot-scale HRAPs with a holistic approach, considering not only the wastewater treatment efficiency and biomass characteristics, but also the bioenergy recovery potential from harvested biomass. In particular, the present study focused on: 1) studying the performance of two parallel pilot systems: a HRAP treating raw urban wastewater and a HRAP treating primary settled urban wastewater; 2) comparing the biomass productivity, composition and settling capacity of each system; and 3) assessing the biogas production potential from microalgal biomass of each system. This is, to the best of the authors knowledge, the first study that explicitly investigated the role of the primary treatment in HRAP systems based on pilot-scale experiments and its effect on bioenergy recovery.

## 2. Materials and methods

### 2.1. High rate algal ponds

Experiments were carried out in a pilot plant located outdoors at the laboratory of the GEMMA Research Group (Universitat Politècnica de Catalunya, Barcelona, Spain) during 260 days (November 2016 – July 2017). The system treated real wastewater from the municipal sewer, which received a pretreatment (screening) in the homogenization tank (1.2 $m^3$) that was continuously stirred to avoid solids sedimentation. From this tank, wastewater was conveyed to two parallel treatment lines: one with a primary treatment (PT) in a cylindrical PVC settling tank (diameter: 18 cm, height: 30 cm, effective volume: 3 L, HRT: 41 min) as a control line (HRAP-PT); and another one without PT as a test line (HRAP-noPT). Subsequently, two identical HRAPs received the corresponding influents (105 L/day) with a HRT of 4.5 days. The HRAPs were made of PVC with a useful volume of 0.47 $m^3$, a surface area of 1.5 $m^2$, a water depth of 0.3 m, and with a paddle wheel constantly stirring the mixed liquor at an average velocity of 10 m/h. Both HRAPs were followed by secondary settlers (diameter: 18 cm, height: 34 cm, effective volume: 3.3 L, HRT: 46 min) where the secondary effluent was separated from the microalgae. The biomass then was further thickened before undergoing anaerobic digestion. Details on the bioenergy recovery set-up will be described later. A schematic structure of the pilot plant is shown in Fig. 1. The performance of both lines were compared in terms of wastewater treatment efficiency and biomass productivity, composition and settling capacity. In order to account for the seasonality, the wastewater treatment efficiency was compared in cold (November to March) and warm (April to July) periods.

**Please insert Figure 1**

*2.2. Wastewater characterisation*

In order to evaluate the wastewater treatment efficiency of both systems, the following parameters were monitored: dissolved oxygen (DO) and temperature (EcoScan DO 6, ThermoFisher Scientific, USA) (daily), pH (Crison 506, Spain) and turbidity (Hanna HI 93703, USA) (three times per week), total suspended solids (TSS), volatile suspended solids (VSS), chlorophyll-a, according to Standard Methods (APHA-AWWA-WEF, 2012), $NH_4^+$-N according to Solórzano method (Solórzano, 1969) and $NO_2^-$-N, $NO_3^-$-N and $PO_4^{3-}$-P through isocratic mode with carbonate-based eluents at a temperature of 30°C and a flow of 1 mL/min (ICS-1000, Dionex Corporation, USA) (limits of detection (LOD) were 0.9 mg/L of $NO_2^-$-N, 1.12 of $NO_3^-$-N, and 0.8 mg/L of $PO_4^{3-}$—P) (twice a week), alkalinity, total and soluble chemical oxygen demand (COD and sCOD) according to Standard Methods (APHA-AWWA-WEF, 2012), total carbon (TC) and total nitrogen (TN) (multi N/C 2100S, Analytik Jena, Germany) (once a week). All the analyses were done in triplicate and results are given as average values.

*2.3. Biomass composition and productivity*

Samples of biomass were analysed microscopically (BA310, Motic, China) once a month, in order to observe the composition of microorganisms and measure flocs sizes during the experimental period. The identification of microalgae genera was based on conventional taxonomic books (Palmer, 1962; Streble and Krauter, 1987).

Average biomass productivity (g VSS/m²d) was calculated based on the VSS concentration in the HRAPs mixed liquor samples, using Equation 1.

$$Biomass\ productivity = \frac{VSS\ (Q - Q_E + Q_P)}{A} \qquad \text{Eq. 1}$$

where $VSS$ is the volatile suspended solids concentration of the HRAP mixed liquor (g VSS/L); $Q$ is the wastewater flow rate (L/d); $Q_E$ is the evaporation rate (L/d); $Q_P$ is the precipitation rate (L/d); and $A$ is the surface area of the HRAP (m2). The evaporation rate was calculated using Eq. 2.

$$Q_E = E_p\ A \qquad \text{Eq. 2}$$

where $A$ is the surface area of the HRAP (m2) and $E_p$ is the potential evaporation (mm/d), calculated from Turc's formula (Eq. 3) (Fisher and Pringle III, 2013).

$$E_p = a\ (R + 50)\ \frac{T_a}{(T_a + 15)} \qquad \text{Eq. 3}$$

where $R$ is the average solar radiation in a day (cal/cm2d); $T_a$ is the average air temperature in a day (°C); and $a$ is a dimensionless coefficient which varies depending on the sampling frequency (0.0133 for daily samples).

Solar radiation, air temperature and precipitation data were provided by the local automatic weather station of Barcelona – Zona Universitària (X8) (Supplementary materials) (DAM, 2017).

*2.4. Biomass settling capacity*

Sedimentation tests were carried out monthly in order to observe the difference between the settling characteristics of the biomass produced in both HRAPs. The tests were performed in a settling column (height: 50 cm, diameter: 9 cm) with four sampling ports at different depths along the column ($d_1$ = 12 cm, $d_2$ = 20 cm, $d_3$ = 32 cm and $d_4$ = 40 cm), according to the method described by Metcalf & Eddy (2003). Mixed liquor of each HRAP was poured into the column up to 45 cm height in such a way that the distribution of particle sizes was uniform from top to bottom. At various time intervals (0, 5, 10, 20, 40, 60, 90, 120, 180 min), samples of 20 mL were withdrawn from the sampling ports and analysed for TSS concentrations. Removal efficiencies were calculated from initial and final TSS concentrations at different time intervals and column depths. Moreover, average settling velocities were estimated considering the column depth and the time needed to reach a certain biomass recovery efficiency.

## 2.5. Biochemical methane potential test

BMP tests were carried out between operational days 213 and 260 in order to compare the biogas production potential of biomass harvested from both systems. BMP tests were performed in serum bottles of 160 mL filled up to 100 mL of liquid volume with certain amounts of inoculum and substrate, corresponding to 5 g VS substrate/L and a substrate to inoculum ratio (S/I) of 0.5 g VS substrate/g VS inoculum (Passos et al., 2013). The substrates used were primary sludge (PS) from the primary settler of the HRAP-PT and microalgal biomass from both the HRAP-PT and HRAP-noPT. PS was purged daily from the primary settler by means of a pump and microalgal biomass was harvested from the secondary settlers following the HRAPs and thickened by gravity in laboratory Imhoff cones at 4°C for 24h (Fig. 1). The microalgae thermal pretreatment was carried out at 75°C for 10h, according to the methodology described by Solé-Bundó et al. (2018).

Microalgal biomass was tested untreated (Microalgae-PT and Microalgae-noPT from the HRAP-PT and HRAP-noPT, respectively) and thermally pre-treated (TPT Microalgae-PT and TPT Microalgae-noPT from the HRAP-PT and HRAP-noPT, respectively). Moreover, in order to increase the C:N ratio, co-digestion (i.e. digestion of a mixture of different substrates) of Microalgae-PT and TPT Microalgae-PT with PS at two different ratios (25% Microalgae - 75% PS and 50% Microalgae - 50% PS on a VS basis) was also tested (Lu and Zhang, 2016). These ratios represent the average volume of microalgae and primary sludge obtained in warm and cold months in a pilot HRAP system (Solé-Bundó et al., 2015). Each trial was performed in triplicate.

After being flushed with helium gas and closed with butyl rubber stoppers, the bottles were placed in a platform shaker incubator (OPAQ, Ovan, Spain) at 35°C and 100 rpm until daily methane production was less than 1% of the total accumulated methane yield in all bottles. Pressure in each bottle was periodically measured with a digital manometer (GMH 3151 Greisinger, Germany) and biogas production was calculated by subtracting the blank (only inoculum) production. The methane content in biogas was analysed by gas chromatography (Trace GC Thermo Finnigan, USA), following the procedure described by Solé-Bundó et al. (2018). The anaerobic biodegradability of each substrate was calculated based on the net methane production (mL $CH_4$) and the theoretical methane yield under standard conditions, which is estimated as 350 mL $CH_4$ for each gram of degraded COD (Chernicharo, 2007).

Microalgal biomass macromolecular composition was expressed in terms of proteins, carbohydrates and lipids over the VS content. Carbohydrates were measured by phenol-sulphuric acid method with acid hydrolysis and determined by spectrophotometry (Spectronic Genesys 8), proteins were measured from the Total Kjeldahl Nitrogen (TKN) (APHA-AWWA-WEF, 2012) and a TKN/protein conversion factor of 5.95 (González López et al., 2010) and lipids were measured with the Soxhlet extraction method, using a mixture of chloroform and methanol at the ratio of 2:1 (v/v) as extractant agents (Folch et al., 1957).

*2.6. Statistical analyses*

Experimental data obtained from the systems HRAP-PT and HRAP-noPT regarding wastewater treatment efficiency, as well as biomass productivity and settleability, were analysed by paired two-sample t-test (α = 0.05) using Minitab 18 (Minitab Inc., PA, USA). For the evaluation of kinetic parameters of the BMP tests, experimental data were adjusted to a first-order kinetic model by the least square method (Schroyen et al., 2014), using the tool *Solver* from Microsoft Excel 2016 (Eq. 8).

$$P = P_o \cdot [1 - \exp(-k \cdot t)] \quad \text{Eq. 8}$$

where $P_o$ stands for the methane production potential (mL $CH_4$/g VS), $k$ is the first order kinetic rate constant (day$_{-1}$), $P$ is the accumulated methane production at time $t$ (mL $CH4$/g VS) and $t$ is time (day).

The error variance ($s^2$) of modelled methane production from Eq. 8 based on the actual methane production was estimated by the following equation (Eq. 9):

$$s^2 = \frac{\sum_1^i (y_i - \hat{y}_i)}{N - K} \quad \text{Eq. 9}$$

where $y_i$ is the experimental value, $\hat{y}_i$ is the value estimated by the model, $N$ is the number of samples and $K$ is the number of model parameters.

The results were statistically assessed via multi-factor analysis of variance (ANOVA) ($\alpha$ = 0.05). The Fisher's Least Significant Difference (LSD) ($\alpha$ = 0.05) was used as a post-hoc test using Minitab 18 (Minitab Inc., PA, USA).

## 3. Results and discussion

### 3.1. Wastewater treatment efficiency

The average values of the main parameters measured in HRAP-PT and HRAP-noPT over a period of 260 days are shown in Table 1 (mixed liquor) and Table 2 (influent and effluent). The temporal variations of water quality parameters monitored in both systems are shown in Figure 2. Moreover, a summary of the average removal efficiencies of the main water quality parameters is shown in Table 3. Additional data on average concentrations and removal efficiencies are presented in Supplementary materials.

The results obtained from the HRAPs indicated that there was no significant difference in terms of wastewater treatment efficiency between the two configurations considered.

**Please insert Table 1**

Average TSS and VSS concentration in the mixed liquor of HRAP-noPT were 41% and 31% significantly higher than in the HRAP-PT, respectively (Table 1). As expected, the difference between the two systems relied more on the higher inert solids concentration discharged into the HRAP-noPT than in microorganisms' biomass (VSS). The average DO concentration in the HRAP-PT was 16% higher compared to the HRAP-noPT (Table 1), which is explained by its lower TSS concentration in the mixed liquor, enhancing light penetration through the pond and leading to a higher photosynthetic activity rate. However, the higher average chlorophyll-a concentration in HRAP-noPT indicates that in spite of the higher solids concentrations, microalgae growth was not hindered in this system.

**Please insert Table 2**

Regarding the wastewater quality parameters, there were no significant differences when comparing $NH_4^+$-N, TN, TC and COD removal efficiencies throughout the entire experimental period between the HRAP-PT and HRAP-noPT (Table 3). Considering the seasonal influence, there were no significant differences in removal efficiencies between the HRAP-PT and HRAP-noPT, except for $NH_4^+$-N and sCOD removal (Table 3). The $NH_4^+$-N removal efficiency was slightly higher in the HRAP-PT during the warm season. This was probably because the proportion of microalgae (as mg chlorophyll-a/g VSS) increased by 61% from cold to warm season in the HRAP-PT, while in the HRAP-noPT the increase was only 6%. The higher microalgae proportion in the HRAP-PT during the warm season could have enhanced the $NH_4^+$-N removal in this system. Similarly, the higher sCOD removal in the HRAP-noPT during the cold season (Table 3) could be related to the higher biomass concentration in this system (Table A.1).

**Please insert Table 3**

Despite the very high removal efficiencies of $NH_4^+$-N (around 90%) in both systems, the TN removal efficiencies were lower (around 45%) (Table 3). This was due to the fact that the influent nitrogen (mainly $NH_4^+$) was converted into $NO_3^-$ (mostly) and $NO_2^-$ (i.e. nitrification), as observed in previous studies (de Godos et al., 2016; Van Den Hende et al., 2016). Moreover, during the warm season photosynthetic activity is enhanced, increasing pH and favouring $NH_4^+$ volatilisation (de Godos et al., 2016; García et al., 2006). This explains the lower $NO_3^-$ effluent concentrations during the warm season compared to the cold season, since a lower amount of $NH_4^+$ was available to be converted into $NO_3^-$ (Figure 2). Average concentrations of $NO_2^-$ in both ponds were very low (up to 2.5 mg/L). Thus, considering also that average $NO_3^-$ concentrations in the influent and effluent of both HRAPs were similar (Figure 2), as well as $NH_4^+$ removal, it can be deduced that the nitrogen conversion pathway was similar in both systems through the experimental period. In general, $NH_4^+$ is the preferential form of nitrogen uptake for most microalgae species, followed by $NO_3^-$ (Maestrini, 1982; Oliver and Ganf, 2002; Ruiz-Marin et al., 2010), which is in accordance with the results obtained in this study.

**Please insert Figure 2**

On the whole, both systems presented high nutrients and organic matter removal efficiencies in spite of the seasonal changes and different operational conditions (i.e. absence of primary treatment). Average COD removal efficiencies ranged between 60 and 67% in both systems through the entire experimental period (Table 3). These removal efficiencies were in accordance with previous studies under similar operational conditions (Young et al., 2017; Sutherland et al., 2014). Another study which evaluated the growth of *Chlorella* sp. in raw and primary treated wastewater from a conventional municipal wastewater plant (i.e.

activated sludge system), also reported similar organic matter and nutrients removal efficiencies (Wang et al., 2010). Average $NH_4^+$-N removal efficiencies were 82.4 and 74.7%, while for COD the removal rates were 50.9 and 56.5% for algae cultivation in wastewater sampled before and after primary treatment, respectively (Wang et al., 2010). Although these results were obtained from batch cultures, the removal efficiencies were similar to the ones found in this work.

The results of this work are in accordance with previous studies in which microalgae were cultivated at lab-scale using wastewater from different stages of municipal wastewater treatment plants, obtaining efficient treatment (Cabanelas et al., 2013; Kong et al., 2009). Furthermore, the present study corroborates with the hypothesis proposed by Posadas et al. (2017) who suggested that, based on a theoretical study, primary suspended solids removal is unlikely needed when using the HRAPs technology for treating urban wastewater.

Finally, based on the results presented in this section, the primary treatment preceding a HRAP seems to be a dispensable step when urban wastewater treatment is the main objective. Moreover, the simplification of a HRAP system by removing the primary treatment step would also incentivise its implementation in small communities, since the wastewater treatment plant footprint and cost could be reduced.

*3.2. Biomass composition and productivity*

Considering the entire experimental period, the HRAP-noPT had a higher biodiversity of microorganisms compared to the HRAP-PT. During the cold season, the microalgal biomass in the HRAP-PT was mainly composed of *Chlorella* sp., while in the HRAP-noPT the predominant microalgae genus was *Stigeoclonium* sp., which formed macroscopic filamentous flocs. However, during the warm season *Chlorella* sp. became the predominant genus in the HRAP-noPT system as well. Diatoms (mostly *Nitzschia* sp. and *Navicula* sp.) and grazers (ciliate and flagellate protozoans) were observed in both ponds along the entire period, but in larger quantity in the HRAP-noPT than in the HRAP-PT (Supplementary materials). The average size for the flocs observed in the HRAP-PT was 50-500 μm, while for the HRAP-noPT it ranged from 100 to 2,000 μm. The biomass diversity is a relevant parameter to be monitored, since it influences downstream processes, such as biogas and bioproducts generation. The presence of grazers, for instance, might affect the productivity of high-value compounds extracted from the biomass.

Microalgal biomass productivity of both HRAPs is shown in Fig. 3. The overall average biomass productivity in the HRAP-noPT was $20 \pm 7$ g VSS/$m_2$d, which was significantly higher (by 30%) than in the HRAP-PT ($15 \pm 6$ g VSS/$m_2$d). Park and Craggs (2010) operated a HRAP with a HRT of 4 days and reported an average biomass productivity of 20.7 g VSS/$m_2$d, which was slightly higher than in the present study most probably because there was $CO_2$ addition to control the pH and prevent carbon limitation. Similar results were described by de Godos et al. (2016), with an average biomass productivity ranging from 13.2 g VSS/$m_2$d (HRT of 5 days in spring) to 23.9 g VSS/$m_2$d (HRT of 3 days in summer) in HRAPs operated without $CO_2$ injection.

The higher biomass productivity observed in the HRAP-noPT might be explained by the higher influent VSS concentration (Table 1). Indeed, the VSS concentration in the influent was 49% higher in the HRAP-noPT than in the HRAP-PT (Table 1). Moreover, the VSS and chlorophyll-a concentrations in the mixed liquor were around 31% and 50% higher in HRAP-noPT than in the HRAP-PT, respectively (Table 1). With this in mind, it can be assumed that part of the VSS introduced in the HRAP-noPT was consumed by the microalgal-bacterial biomass. In other words, the VSS in the influent (i.e. organic matter from the wastewater) was converted into microalgal-bacterial biomass in the HRAP-noPT system, where the microalgal proportion may have increased better than in the HRAP-PT system. As mentioned before, the difference in TSS influent concentration (Table 1) and, consequently, on the light availability between the two systems, did not seem to have created photo-inhibition. Indeed, previous studies, which investigated the composition of the phytoplankton community in three HRAPs submitted to different solar radiation levels, also reported that light availability was not the main influence on the growth and development of microalgal biomass. Other aspects, such as competition with other microorganisms for space and nutrients, and predation by zooplankton seemed to have a higher effect on microalgae biomass composition and productivity (Assemany et al., 2015).

With regards to seasonal influence, there was a slight increase in biomass productivity in warmer months (Figure 3). It is worth noting that during those months, the abundance of grazers in both ponds also increased. The presence of these predators indicated that the actual biomass productivity might have been higher that the calculated values, which were based on the VSS concentrations measured in the mixed liquor of both ponds. This could possibly explain the high variation seen in June, in which the ranges of biomass productivity measured in both ponds were the largest of the entire period (HRAP-PT: 5 - 33 gVSS/m$_2$d and HRAP-noPT: 14 - 46 gVSS/m$_2$d). Biomass losses caused by these organisms have also been reported in previous studies (Mehrabadi et al., 2016; Montemezzani et al., 2016; Park et al., 2013). Finally, although the HRAP-noPT received higher organic loading, the production of microalgal biomass was not jeopardised. In addition, the higher biomass productivity would most likely lead to higher biogas production per day or other bioproducts obtained from this biomass.

**Please insert Figure 3**

*3.3. Biomass settling capacity*

The biomass sedimentation through gravity settling was assessed by monthly settling column tests. The assessment of the settling capacity helps to define further harvesting and dewatering techniques to be applied at large scale, which usually represents high energy consumption on the overall process (Fasaei et al., 2018). In this study, the initial biomass concentration in the mixed liquor varied from 0.26 – 0.39 g VSS/L for the HRAP-PT and 0.23 – 0.72 g VSS/L for the HRAP-noPT. As mentioned above, biomass recovery efficiencies were calculated from the initial and final TSS concentrations at different time intervals and column depths.

The settling tests results indicated that the biomass from both systems had good settling capacity. Figure 4a shows the biomass recovery over time with curves representing the four different sampling depths (12, 20, 32 and 40 cm). Based on these data, the time required to obtain certain biomass recovery efficiencies (80, 85, 90 and 95%) was calculated (Fig. 4b). Considering average values of all settling tests, the biomass from the HRAP-noPT was faster to reach recovery efficiencies of 80, 85 and 90%, and the HRAP-PT was faster only for 95% recovery. This is in accordance with microbiology observations, that recorded higher biodiversity of microorganisms for the HRAP-noPT than the HRAP-PT during the entire period. Moreover, filamentous microalgae present in the HRAP-noPT during the cold season, which are organisms linked to flocs aggregation, also influenced the higher settling capacity of this biomass.

**Please insert Figure 4**

Biomass recovery efficiencies were lower than those found in a previous study with similar biomass composition, with about 85% recovery in less than 40 min (Gutiérrez et al., 2015). However, it is important to mention that the initial biomass concentration in that study was higher (800 mg VSS/L) than in the present one (300 - 400 mg VSS/L). In that study, the average time needed to recover 90% of biomass was 58 min, with a final effluent concentration of 80 mg VS/L. In the present study, the average times needed to reach 90% of biomass recovery was 129 min (HRAP-PT) and 114 min (HRAP-noPT), but the final effluent concentrations were much lower: 30 and 40 mg VSS/L. This highlights the importance of considering the final effluent quality when comparing results of relative removal efficiencies from different studies.

The relation between the sampling depth and settling time recorded for biomass from the HRAP-PT and HRAP-noPT is illustrated by isorecovery curves (Fig. 4b). Each curve shows the time required to obtain a certain biomass recovery at different depths. Thus, the settling velocities were calculated by dividing the column depth ($d_i$) by time ($t_i$).

For instance, the average settling velocities for 80% recovery were 0.47 and 0.51 m/h, and for 95% recovery they were 0.13 and 0.09 m/h for the HRAP-PT and HRAP-noPT, respectively. For 80% recovery, the HRAP-noPT had a slightly higher velocity, which is explained by the larger flocs, but for 95% HRAP-PT had a higher velocity, indicating the higher amount of colloidal particles in the HRAP-noPT resulting from the influent characteristics. The settling velocities were similar to the ones reported by Moorthy et al. (2017), which ranged from 0.03 to 0.08 m/h for *Scenedesmus abundans*, and by Peperzak et al. (2003), which fluctuated from 0.02 to 0.09 m/h for a mixture of microalgae.

Overall, the biomass from both systems presented good settling capacity with no significant differences between them. Thus, the absence of primary treatment did not affect the biomass settling capacity.

*3.4. Biochemical methane potential test*

The BMP test was performed in order to complement the comparison between the HRAP-PT and HRAP-noPT, in terms of potential bioenergy recovery from biomass harvested in each system. Biochemical analysis indicated that microalgal biomass was mainly composed of proteins (41 - 49%), followed by carbohydrates (27 - 33%) and lipids (20 - 25%) (Table 4), in accordance with previous studies (Dong et al., 2016; Solé-Bundó et al., 2017a).

**Please insert Table 4**

The methane yield of each trial over an incubation period of 48 days is illustrated in Fig. 5. The methane content in biogas was similar in all cases (around 72%).

The lowest methane yield was obtained in the mono-digestion of Microalgae-noPT, with a final yield of 188.7 mL $CH_4$/g VS; and the highest methane yield was from the co-digestion of 25% Microalgae-PT + 75% PS, reaching a final yield of 258.3 mL $CH_4$/g VS. This was 25% higher compared to the mono-digestion of the Microalgae-PT. During the initial stage of the incubation (especially the first 6 days) the kinetics and productions were better for TPT

Microalgae-PT, TPT Microalgae-noPT and Microalgae-noPT (Fig. 5a). However, after the 9th day the behaviour changed and the Microalgae-PT production slightly increased compared to Microalgae-noPT (both untreated and TPT). This performance could be explained by the fact that Microalgae-noPT contained more readily biodegradable material (which was transformed into biogas) than the Microalgae-PT, as expected, since the former was harvested from the system without primary treatment.

**Please insert Figure 5**

The final methane yield of pre-treated microalgae from the HRAP-PT, primary sludge and its co-digestion with untreated or pre-treated microalgae grown in the HRAP-PT were not statistically different from each other (Table 5). In addition, no significant differences were found in the final methane yield from untreated and pre-treated microalgae grown in both HRAP-noPT and HRAP-PT (Table 5). Nevertheless, the methane yield of untreated and pre-treated microalgae grown in HRAP-noPT were significantly lower than those obtained with the co-digestion of primary sludge and microalgae harvested in the HRAP-PT (Table 5).

**Please insert Table 5**

The thermal pretreatment was applied in this study in order to increase microalgae biodegradability by breaking down their resistant cell wall, as suggested by previous studies (Solé-Bundó et al., 2018). Several studies on microalgae pretreatment for biogas production have been reported, including biological, chemical and physical pretreatments (Kendir and Ugurlu, 2018). The selection of a thermal pretreatment for this study was based on previous research comparing different pretreatments, which showed that the thermal one would reach the highest methane yield and considerably better energy balance (Kendir and Ugurlu, 2018; Passos et al., 2015). Comparing the mono-digestions, the thermal pretreatment improved the methane yield by 3% (HRAP-noPT) and 9% (HRAP-PT). Although no statistical difference (P-values: 0.80 for HRAP-noPT and 0.37 for HRAP-PT) was found between the methane yield of untreated and thermally pre-treated microalgae from both systems (Table 5), the thermal pretreatment did improve the kinetics in all cases (by 14-22%) as compared to untreated microalgae, which is in agreement with Solé-Bundó et al. (2017c).

In contrast, the co-digestion of microalgae and sludge showed a more significant improvement, increasing the methane yield up to 25% and the kinetics up to 39% compared to microalgae mono-digestion. Moreover, the kinetics of co-digestion with thermally pre-treated microalgae at both ratios (25-75% and 50-50%) were even higher than primary sludge (Table 5). This highlights the synergy of co-digesting microalgae with primary sludge, as also described in previous studies on co-digestion of microalgae and other C-rich substrates (Solé-Bundó et al., 2017b; Yen and Brune, 2007). The results are also in agreement with previous studies in which the co-digestion of microalgae and sewage sludge had a synergistic effect (Olsson et al., 2014; Solé-Bundó et al., 2018).

## 4. Conclusions

The removal of the primary treatment preceding a HRAP, which would simplify its maintenance, reduce costs and the footprint, did not significantly affect the wastewater treatment efficiency. Thus, it seems to be a dispensable step when urban wastewater treatment is the main objective. Although the HRAP without primary treatment received higher organic loading due to the absence of primary treatment, the production of microalgal biomass was not jeopardised. Bioenergy recovery through biogas production would be a good alternative for biomass valorisation. In particular, the co-digestion with primary sludge could improve the methane yield and kinetics of microalgae mono-digestion.

**Acknowledgements**

This research was funded by the Spanish Ministry of Economy and Competitiveness (FOTOBIOGAS Project CTQ2014-57293-C3-3-R) and the European Union's Horizon 2020 research and innovation programme under the Marie Skłodowska-Curie grant agreement No 676070 (SuPER-W). This communication reflects only the author's view and the Research Executive Agency of the EU is not responsible for any use that may be made of the information it contains. Marianna Garfí is grateful to the Spanish Ministry of Economy and Competitiveness (Plan Estatal de Investigación Científica y Técnica y de Innovación 2013–2016, Subprograma Ramón y Cajal (RYC) 2016, RYC-2016-20059). The authors acknowledge Jose Luis Pellín Moreno, Manuel Ángel Arenas Vallejo, Valentin Lebot and Pol Puigseslloses i Sánchez for their contribution to this work.

**Appendix A. Supplementary data**

E-supplementary data of this work can be found in online version of the paper.

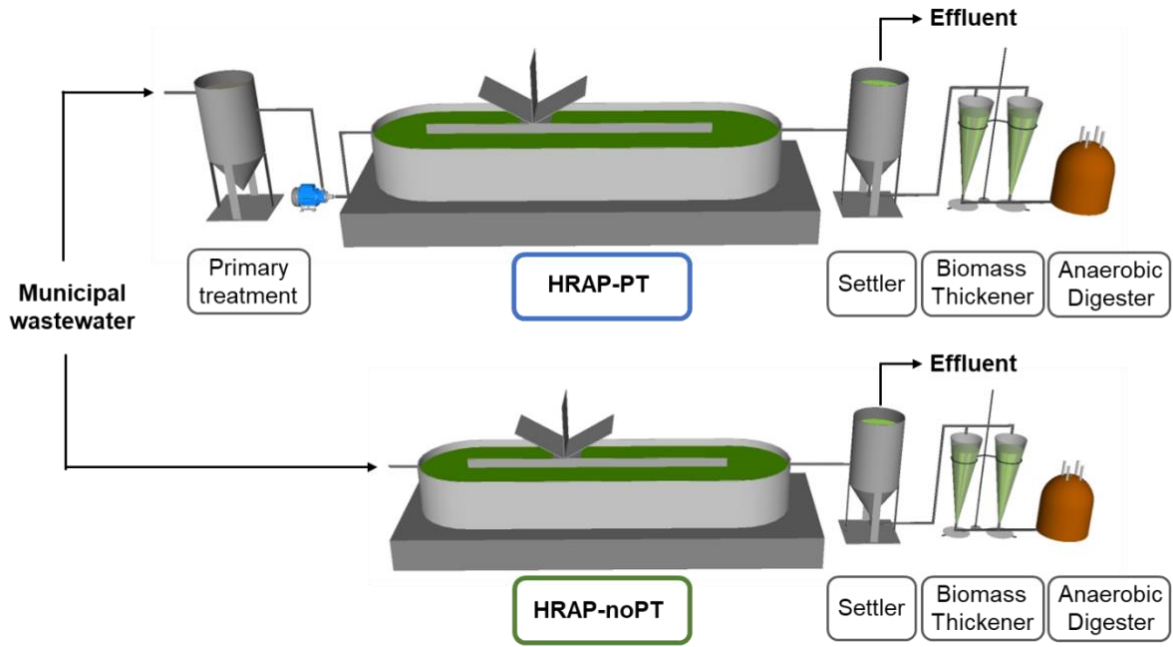

**Figure 1**. Scheme of the microalgae-based wastewater treatment pilot plant located outdoors in Barcelona (Spain). HRAP-PT is the line with primary treatment (PT) and HRAP-noPT is the line without PT.

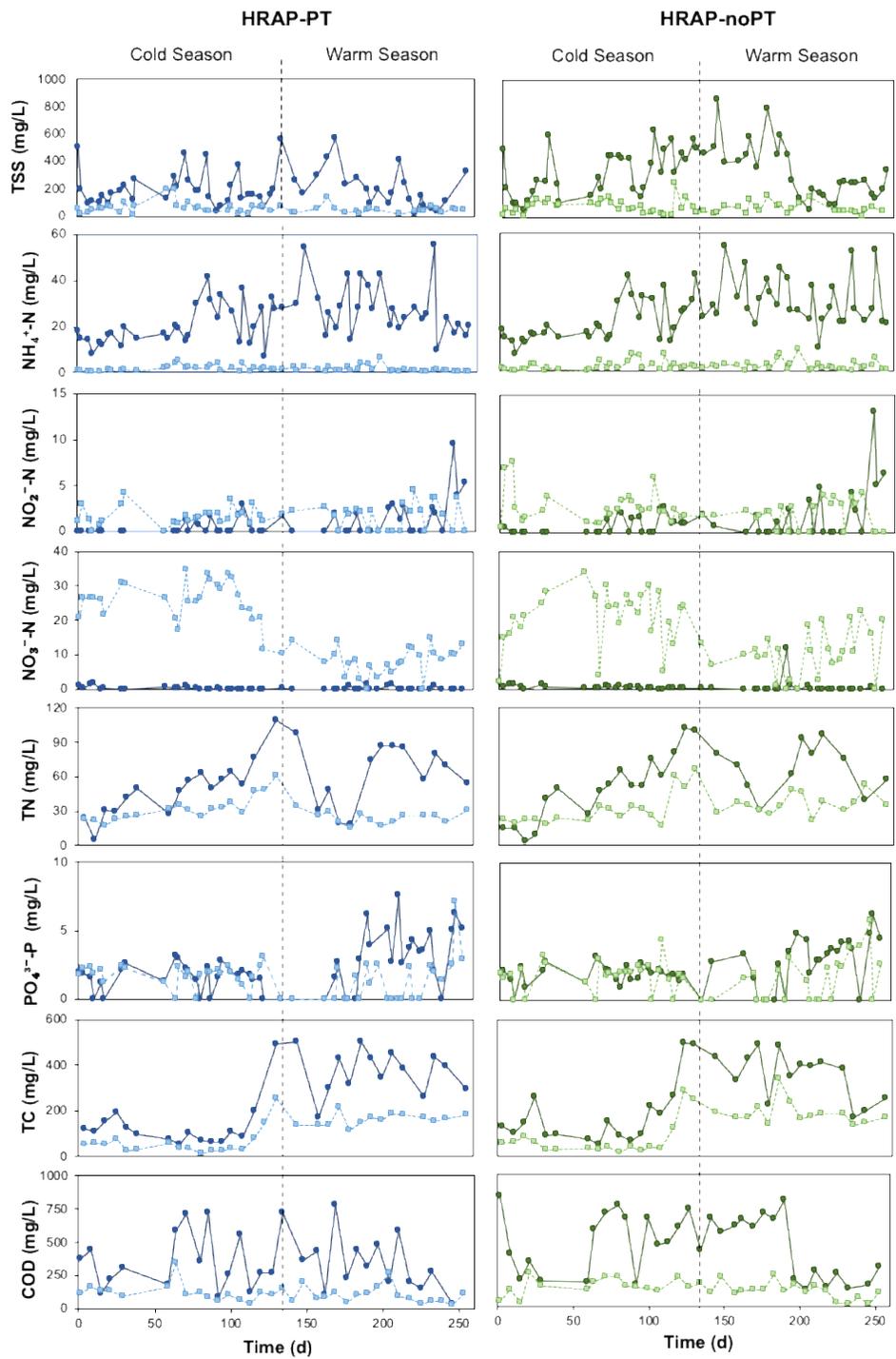

**Figure 2**. Influent (●) and effluent (■) concentrations of total suspended solids (TSS), $NH_4^+$-N, $NO_2^-$-N, $NO_3^-$-N, total nitrogen (TN), $PO_4^{3-}$-P, total carbon (TC) and chemical oxygen demand (COD) measured in the HRAP-PT and HRAP-noPT during the experimental period.

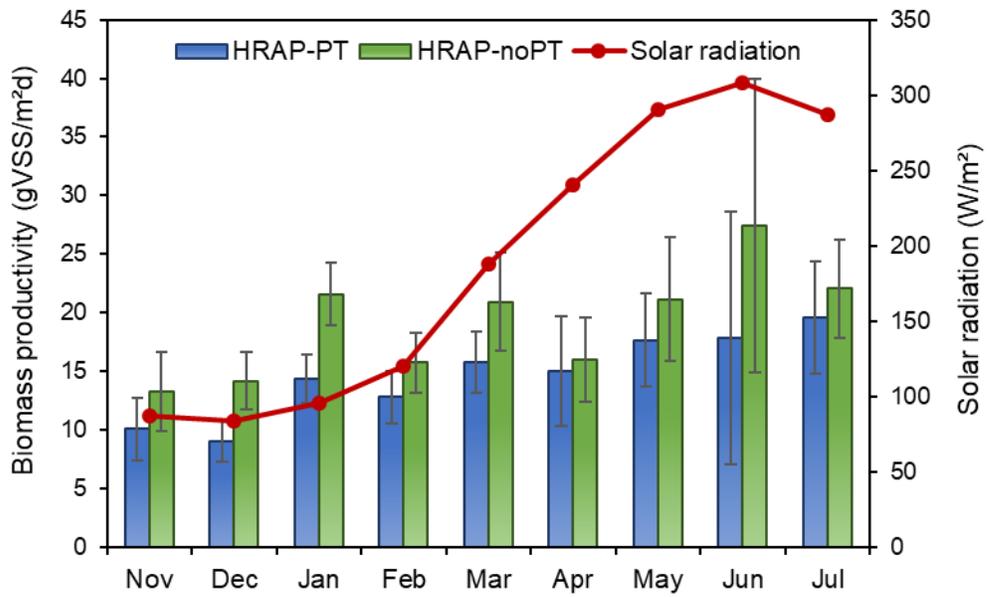

**Figure 3**. Monthly average biomass productivity in the HRAP-PT and HRAP-noPT from November 2016 to July 2017.

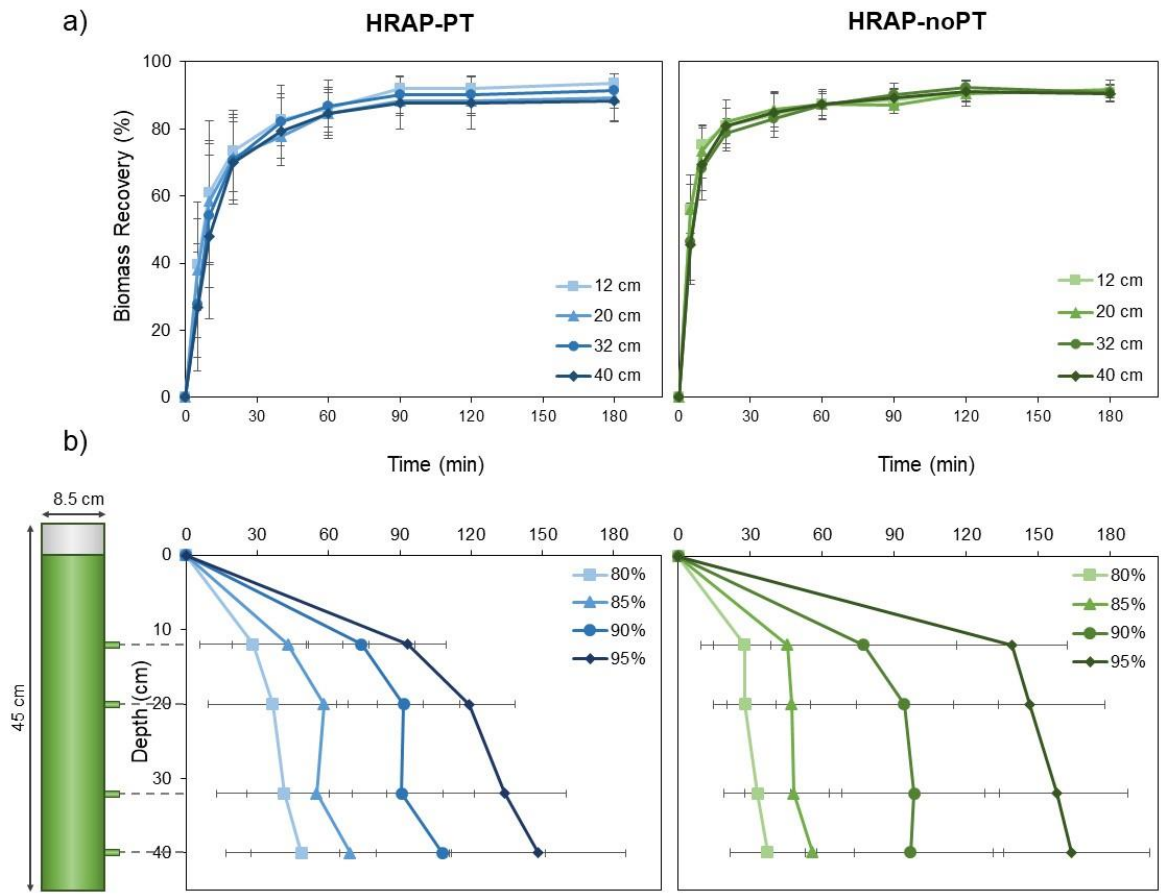

**Figure 4**. Average results of settling tests (n=8) for the HRAP-PT and HRAP-noPT: a) Removal efficiencies at depths of 12 cm (■), 20 cm (▲), 32 cm (●) and 40 cm (♦); b) Average microalgal biomass isorecovery curves of 80% (■), 85% (▲), 90% (●) and 95% (♦).

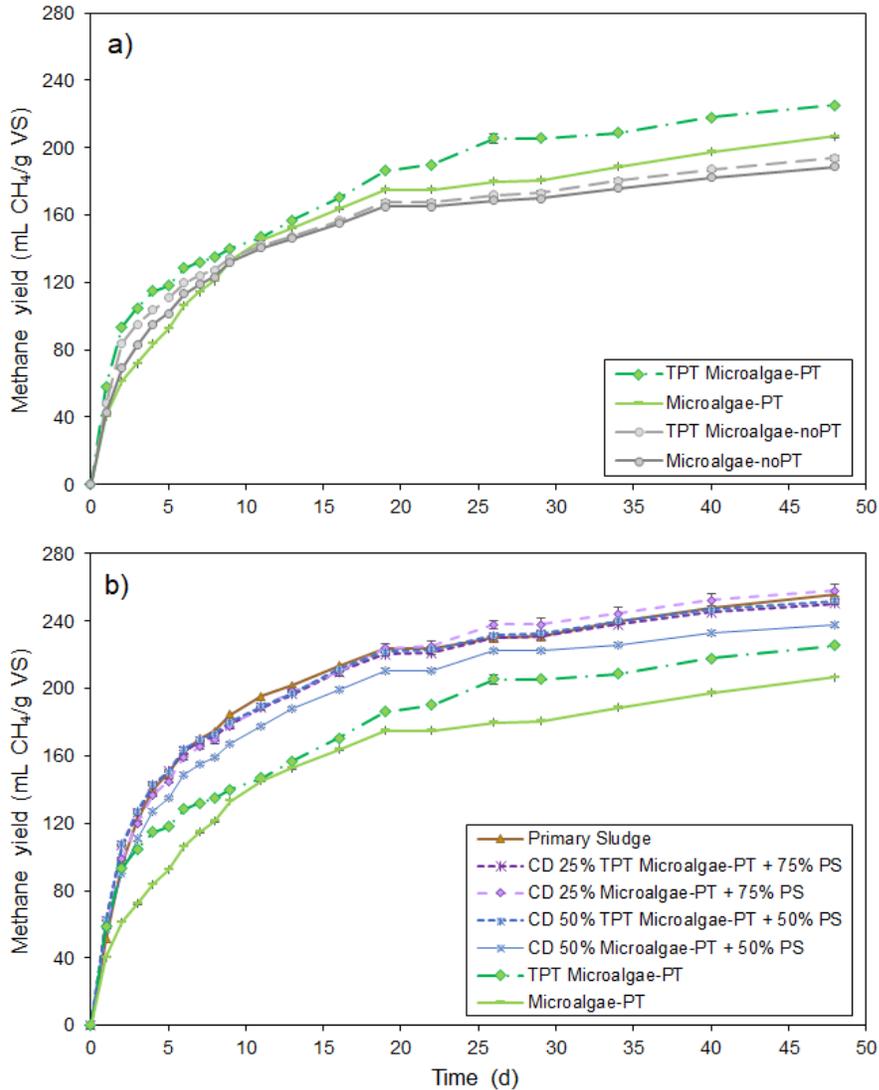

**Figure 5.** Cumulative methane yields showing the effects of: a) thermal pretreatment (TPT), with the comparative results for microalgal biomass from the HRAP-PT and HRAP-noPT: untreated (Microalgae-PT and Microalgae-noPT) and thermally pre-treated (TPT Microalgae-PT and TPT Microalgae-noPT); and b) co-digestion (CD), with the comparative results for Primary Sludge (PS) and co-digestion of Microalgae-PT and TPT Microalgae-PT with PS at two different ratios (25% microalgae + 75% PS and 50% microalgae + 50% PS on a VS basis).

**Table 1**. Summary of the average values of the main parameters monitored in the mixed liquor of both HRAPs through the entire experimental period (260 days). *P*-values for the t-test comparing values of the mixed liquor (95% confidence interval) are highlighted in bold when there is significant difference.

|  | **HRAP-PT** | **HRAP-noPT** | *P*-value |
|---|---|---|---|
| TSS (mg/L) | 261 ± 106 | 370 ± 131 | **9.7E-15** |
| VSS (mg/L) | 230 ± 91 | 301 ± 112 | **1.7E-10** |
| pH | 8.2 ± 0.5 | 7.9 ± 0.3 | **2.1E-13** |
| Turbidity (NTU) | 136 ± 73 | 160 ± 74 | **4.7E-04** |
| TN (mg/L) | 47 ± 13 | 52 ± 15 | **1.5E-02** |
| TC (mg/L) | 226 ± 154 | 240 ± 144 | **2.1E-02** |
| DO (mg/L) | 8.7 ± 2.2 | 7.6 ± 2.2 | **7.8E-20** |
| Chlorophyll-a (mg/L) | 1.1 ± 0.8 | 1.7 ± 0.8 | **1.8E-06** |

Acronyms: TSS (total suspended solids); VSS (volatile suspended solids); TN (total nitrogen); TC (total carbon); DO (dissolved oxygen).

**Table 2**. Summary of the average values of the main parameters monitored in the influent and effluent of both HRAPs through the entire experimental period (260 days).

|  | HRAP-PT | | HRAP-noPT | |
| --- | --- | --- | --- | --- |
|  | Influent | Effluent | Influent | Effluent |
| TSS (mg/L) | 201 ± 132 | 52 ± 37 | 333 ± 183 | 75 ± 46 |
| VSS (mg/L) | 185 ± 112 | 49 ± 32 | 280 ± 143 | 67 ± 38 |
| pH | 7.8 ± 0.3 | 8.0 ± 0.4 | 8.0 ± 0.2 | 7.7 ± 0.2 |
| Turbidity (NTU) | 135 ± 115 | 25 ± 22 | 170 ± 104 | 41 ± 37 |
| TN (mg/L) | 53 ± 27 | 28 ± 10 | 56 ± 28 | 33 ± 12 |
| TC (mg/L) | 244 ± 157 | 107 ± 69 | 258 ± 149 | 126 ± 88 |
| $NH_4^+$-N (mg/L) | 24 ± 11 | 1.5 ± 1.3 | 26 ± 11 | 2.2 ± 2.1 |
| $NO_3^-$-N (mg/L) | 0.2 ± 0.4 | 17 ± 10 | 0.6 ± 1.7 | 16 ± 9 |
| $NO_2^-$-N (mg/L) | 0.9 ± 1.7 | 1.6 ± 1.2 | 1.2 ± 2.3 | 2.3 ± 1.7 |
| $PO_4^{3-}$-P (mg/L) | 2.3 ± 1.8 | 1.5 ± 1.3 | 2.3 ± 1.5 | 1.7 ± 1.4 |
| COD (mg/L) | 353 ± 208 | 114 ± 65 | 464 ± 234 | 134 ± 64 |
| sCOD (mg/L) | 88 ± 48 | 58 ± 31 | 97 ± 47 | 61 ± 38 |

Acronyms: TSS (total suspended solids); VSS (volatile suspended solids); TN (total nitrogen); TC (total carbon); COD (chemical oxygen demand); sCOD (soluble chemical oxygen demand).

Table 3. Summary of the average removal efficiencies of the main water quality parameters measured in the influent and effluent of both HRAPs in cold (Nov-Mar) and warm (Apr-Jul) seasons. *P*-values for the t-test comparing values of the removal efficiencies (95% confidence interval) are highlighted in bold when there is significant difference.

| | Cold Season | | | Warm Season | | | Entire experimental period | | |
|---|---|---|---|---|---|---|---|---|---|
| | Removal (%) | | *P*-value | Removal (%) | | *P*-value | Removal (%) | | *P*-value |
| | HRAP-PT | HRAP-noPT | | HRAP-PT | HRAP-noPT | | HRAP-PT | HRAP-noPT | |
| $NH_4^+$-N | 91 ± 7 | 91 ± 7 | 0.75 | 95 ± 4 | 92 ± 9 | **0.01** | 93 ± 6 | 91 ± 8 | 0.05 |
| TN | 43 ± 9 | 46 ± 16 | 0.37 | 57 ± 21 | 50 ± 17 | 0.34 | 49 ± 17 | 48 ± 16 | 0.73 |
| TC | 59 ± 15 | 61 ± 15 | 0.55 | 54 ± 15 | 44 ± 14 | 0.15 | 56 ± 15 | 53 ± 17 | 0.37 |
| $PO_4^{3-}$-P | 12 ± 47 | 4 ± 55 | 0.66 | 68 ± 38 | 56 ± 44 | 0.19 | 37 ± 52 | 25 ± 52 | 0.22 |
| COD | 60 ± 22 | 63 ± 23 | 0.59 | 64 ± 23 | 67 ± 25 | 0.75 | 62 ± 22 | 65 ± 23 | 0.58 |
| sCOD | 44 ± 19 | 56 ± 22 | **0.03** | 33 ± 18 | 35 ± 16 | 0.77 | 39 ± 19 | 47 ± 22 | 0.08 |

Acronyms: TN (total nitrogen); TC (total carbon); COD (chemical oxygen demand); sCOD (soluble chemical oxygen demand).

5   **Table 4**. Average biochemical composition of the inoculum and substrates used for the BMP
6   test. Microalgae-PT and Microalgae-noPT refer to microalgal biomass harvested from the
7   HRAP-PT and HRAP-noPT, respectively; untreated or thermally pre-treated (TPT).

| Parameter | Inoculum | Primary Sludge | Microalgae-PT | | Microalgae-noPT | |
|---|---|---|---|---|---|---|
| | | | Untreated | TPT | Untreated | TPT |
| pH | 7.35 | 6.37 | 6.46 | 6.74 | 6.33 | 6.48 |
| TS [%(w/w)] | 2.12 ± 0.01 | 3.13 ± 0.04 | 6.09 ± 0.01 | 6.03 ± 0.01 | 5.87 ± 0.02 | 5.80 ± 0.01 |
| VS [%(w/w)] | 1.31 ± 0.13 | 2.32 ± 0.40 | 4.65 ± 0.23 | 4.62 ± 0.28 | 3.96 ± 0.62 | 4.02 ± 0.11 |
| COD (g $O_2$/L) | 16.90 ± 0.50 | 15.43 ± 0.29 | 79.87 ± 0.88 | 79.70 ± 0.25 | 59.43 ± 1.07 | 59.87 ± 1.38 |
| Carbohydrates (%VS) | - | - | 29.7 | 26.9 | 29.0 | 32.5 |
| Proteins (%VS) | - | - | 48.8 | 47.4 | 43.6 | 41.2 |
| Lipids (%VS) | - | - | 20.6 | 25.0 | 22.0 | 19.8 |

Acronyms: TS (total solids); VS (volatile solids); COD (chemical oxygen demand).

9   **Table 5**. Summary of the methane yield (initial after 6 days and final after 48 days of
10   digestion), anaerobic biodegradability (mean values ± standard deviation; n=3) and first-
11   order kinetics constant ($k$) obtained from Eq. 8 (error variance ($s^2$) from Eq. 9 is represented
12   in brackets).

| Substrate | Initial methane yield (mL CH$_4$/g VS d) | | Final methane yield (mL CH$_4$/g VS) | | Anaerobic Biodegradability (%) | | First-order kinetics constant (day$^{-1}$) | |
|---|---|---|---|---|---|---|---|---|
| | **Untreated** | **TPT** | **Untreated** | **TPT** | **Untreated** | **TPT** | **Untreated** | **TPT** |
| Primary sludge | 163.1$_a$ ± 1.1 | | 255.5$_a$ ± 2.4 | | 37.7 ± 2.4 | | 0.202 (135) | |
| Microalgae-noPT | 113.1$_a$ ± 0.4 | 119.8$_a$ ± 0.6 | 188.7$_b$ ± 0.7 | 193.9$_b$ ± 1.4 | 25.3 ± 0.7 | 25.8 ± 1.4 | 0.179 (78) | 0.205 (150) |
| Microalgae-PT | 106.3$_a$ ± 0.2 | 128.5$_a$ ± 0.2 | 206.8$_b$ ± 0.7 | 225.4$_{a,b}$ ± 0.7 | 25.3 ± 0.7 | 26.5 ± 0.7 | 0.135 (63) | 0.165 (326) |
| CD 25% Microalgae-PT + 75% PS | 159.5$_a$ ± 1.7 | 163.1$_a$ ± 0.3 | 258.3$_a$ ± 3.9 | 250.3$_a$ ± 0.4 | 35.1 ± 3.9 | 34.5 ± 0.4 | 0.184 (201) | 0.214 (208) |
| CD 50% Microalgae-PT + 50% PS | 148.8$_a$ ± 1.1 | 164.0$_a$ ± 0.5 | 237.6$_{a,b}$ ± 1.7 | 251.9$_a$ ± 0.5 | 31.1 ± 1.7 | 32.2 ± 0.5 | 0.187 (146) | 0.213 (216) |

13   $_{a,b}$ : Letters indicate a significant difference of methane yield between trials (α = 0.05) after Fisher's LSD test.